# Dual-Comb Spectrometry Enabled by Gain-Assisted Electro-Optic Fiber Ring Resonators

Yuan Chen[1†], Jieming Zhang[1†], Zijian Wang[1,2], Jia Xu[1], Mei Yang[1], Wending Huang[1], Rui Zhu[1], Zhen Hu[1], Xiangze Ma[1], Xiuxiu Zhang[1], Cong Jiang[1], Zhaoyang Wen[1,2], Zhuoren Wan[1,2*], Ming Yan[1,2,3*], and Heping Zeng[1,2,4]

[1]State Key Laboratory of Precision Spectroscopy, and Hainan Institute, East China Normal University, Shanghai 200062, China

[2]Chongqing Key Laboratory of Precision Optics, Chongqing Institute of East China Normal University, Chongqing 401120, China

[3]Collaborative Innovation Center of Extreme Optics, Shanxi University, Taiyuan, Shanxi 030006, People's Republic of China

[4]Jinan Institute of Quantum Technology, Jinan, Shandong 250101, China

† These authors contribute equally.

* Corresponding authors: *zrwan@lps.ecnu.edu.cn; myan@lps.ecnu.edu.cn*.

**Abstract**

Broadband dual-comb spectroscopy (DCS) with widely tunable repetition rates and simplified frequency stabilization is desirable for diverse sensing applications. Here, we demonstrate DCS using two gain-assisted fiber ring resonators that share a continuous-wave (CW) laser and incorporate intracavity electro-optic (EO) phase modulation. Intracavity gain compensates for round-trip losses, sustaining repeated phase modulation and enabling flat-top comb spectra with a 20-dB bandwidth exceeding 1.8 THz. Harmonic-order selection provides widely tunable repetition rates from approximately 5 to 0.2 GHz without substantially compromising the spectral bandwidth. The comb phase noise closely follows that of the radio-frequency (RF) drive at low offset frequencies. Without active frequency stabilization, the two combs exhibit an instrument-limited beat-note linewidth of 2 Hz, indicating high mutual coherence. The generated dual combs enable high-speed DCS at the single-interferogram rate, providing a platform for dynamic spectroscopic measurements. This all-fiber architecture, together with its simplified operation, offers a promising route towards compact and readily integrable dual-comb instruments.

## 1. Introduction

Optical frequency combs generate broadband spectra consisting of evenly spaced spectral lines, revolutionizing molecular spectroscopy [1]. In particular, dual-comb spectroscopy (DCS), which relies on the multiheterodyne beating of two optical frequency combs with slightly different repetition rates, enables rapid, broadband and high-resolution measurements for a wide range of spectroscopic sensing applications [2, 3]. In a conventional dual-comb spectrometer [4], the finest single-shot spectral resolution is determined by the comb repetition rate ($f_{\mathrm{rep}}$), which sets the comb-line spacing; the measurement refresh rate is governed by the repetition-rate difference ($\Delta f_{\mathrm{rep}}$); and resolving individual multiheterodyne beat notes requires mutual coherence between the two combs [5]. A single dual-comb platform offering high mutual coherence and widely tunable repetition rates for diverse sensing applications is highly desirable [6], yet remains challenging to realize.

Mode-locked laser combs with repetition rates of a few hundred megahertz (MHz) have been workhorses for broadband, high-resolution DCS, enabling applications such as precision molecular spectroscopy [7], trace-gas detection [8] and human breath analysis [9]. Their relatively low repetition rates also provide pulse energies exceeding 1 nJ and femtosecond pulse durations, making them suitable for nonlinear spectroscopic measurements, including two-photon absorption spectroscopy [10] and coherent Raman spectroscopy [11]. More recently, laser combs with repetition rates above 1 GHz have attracted interest for atmospheric gas sensing [12] and combustion diagnostics [13], where molecular absorption linewidths typically exceed several gigahertz (GHz). In these applications, repetition rates above 1 GHz offer a favorable trade-off between spectral resolution and measurement refresh rate, facilitating the capture of dynamic processes [13]. However, the repetition rates of mode-locked combs are fundamentally constrained by their cavity round-trip times, making wide-range tuning of the comb-line spacing challenging [14]. Furthermore, maintaining mutual coherence between two independent mode-locked lasers typically requires stabilization or post-processing correction of their relative timing and optical phase, adding to system complexity [15, 16].

Alternatively, electro-optic (EO) comb generation offers widely tunable repetition rates and excellent mutual coherence [17-19]. This flexibility stems from the fact that the comb repetition rate is directly determined by the radio-frequency (RF) driving signal, while the optical carrier frequency is inherited from a continuous-wave (CW) seed laser. However, the bandwidth of directly generated EO combs is fundamentally limited by the available modulation depth. Several approaches have been developed to overcome this limitation. Cascaded modulation, which combines intensity and phase modulators (PMs) or employs tailored microwave waveforms, can effectively increase the modulation depth and improve spectral flatness. However, these approaches require additional modulators and microwave driving electronics [20, 21]. An alternative strategy is resonant enhancement, whereby repeated cavity round trips enable the accumulation of higher-order sidebands, effectively increasing the modulation depth beyond that achievable in a single pass [22, 23]. This approach has enabled broadband EO comb generation in integrated photonic platforms [24, 25]. Nevertheless, in passive resonators, round-trip losses and dispersion-induced phase mismatch can impede coherent sideband accumulation. Furthermore, the achievable comb bandwidth decreases significantly as the comb repetition rate is reduced (see Methods).

In this paper, we demonstrate DCS using gain-assisted fiber ring resonators driven by a common CW laser. The recirculating architecture incorporates intracavity amplification to compensate for round-trip losses [26-29], thereby sustaining repeated modulation and enabling broadband comb generation. By selecting different harmonic orders of the cavity free spectral range, the comb repetition rate can be tuned from 5 to 0.2 GHz without substantially compromising the spectral bandwidth. Furthermore, seeding both resonators with the same CW laser, as demonstrated in previous EO dual-comb systems [23], ensures excellent mutual coherence between the two combs. This enables instrument-limited comb linewidths of approximately 2 Hz without active frequency stabilization.

## 2. Methods

Figure 1a illustrates the gain-assisted fiber ring DCS. Each ring contains erbium-doped gain fiber and a PM. On each round trip, the PM generates sidebands around the circulating optical lines. These sidebands add coherently when the injected carrier coincides with a cavity eigenmode and the repetition rate satisfies $f_{\text{rep}} = N \times \text{FSR}$. Here, $N$ is a positive integer and $\text{FSR} = c/(n_{\text{g}}\, L)$ is the cavity free spectral range, where $c$ is the speed of light in vacuum, $n_{\text{g}}$ is the group index, and $L$ is the cavity length. In DCS, the resonant repetition rates satisfy $f_{\text{rep},i} \approx N_i\ \text{FSR}_i$, where $i = 1, 2$ and $N_i$ is the corresponding harmonic order. Their repetition-rate difference is therefore $\Delta f_{\text{rep}} \approx |N_2\ \text{FSR}_2 - N_1\ \text{FSR}_1|$.

For the conventional passive Fabry–Pérot EO comb illustrated in Figure 1b, the power $P_{\text{k}}$ of the $k$th-order line has an approximately exponential envelope, $P_{\text{k}} \propto \exp[-|k|\pi/(\beta \text{F})]$, where $\beta$ is the single-pass phase-modulation index and $F$ is the cavity finesse [22]. With $\kappa$ denoting the full width at half maximum (FWHM) of the cavity resonance in hertz, $F = \text{FSR}/\kappa$. The characteristic total number of comb lines scales as $N_{\text{line}} \approx 2\beta \text{F}/\pi$. At fundamental modulation resonance, $f_{\text{rep}} = \text{FSR}$, the characteristic spectral bandwidth is therefore approximated by

$$\Delta v = N_{\text{line}} f_{\text{rep}} \approx \frac{2\beta}{\pi} F f_{\text{rep}} = \frac{2\beta}{\pi} \frac{f_{\text{rep}}^{2}}{\kappa} \tag{1}$$

For passive reference resonators with the same linewidth $\kappa$ and modulation index $\beta$, Equation (1) gives a quadratic dependence on the fundamental repetition rate as the FSR is varied. In the present architecture, intracavity erbium gain compensates round-trip loss (Figure 1c), sustaining the coherent cascade that produces comb lines separated by the drive frequency [26]. Gain compensation increases the effective number of coherent modulation passes, allowing a broad, flat-top spectrum to be retained at low repetition rates (Figure 1d).

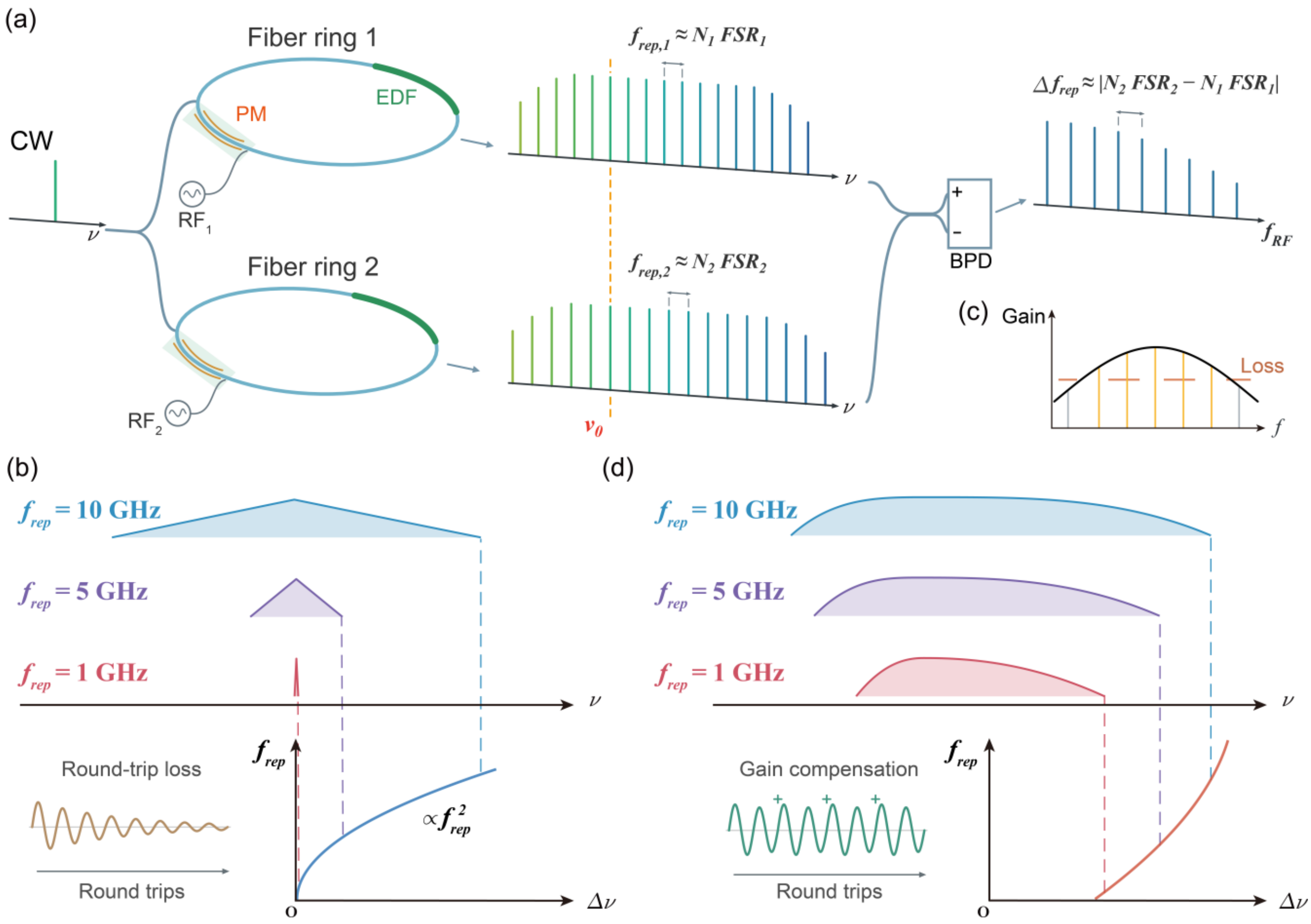

**Figure 1. Gain-assisted electro-optic (EO) dual-comb generation.** (a) Schematic of two fiber ring resonators sharing a continuous-wave (CW) laser, driven by independent radio-frequency (RF) signals and followed by balanced photodetection. (b) Spectral envelopes of passive Fabry–Pérot (FP) resonators and their bandwidth dependence on the repetition rate. (c) Gain spectrum and round-trip loss. (d) Spectral envelopes of gain-assisted resonators and their bandwidth dependence on the repetition rate. Insets in (b) and (d) illustrate round-trip attenuation and gain compensation, respectively. EDF, erbium-doped fiber; PM, phase modulator; BPD, balanced photodetector. For resonator $i = 1, 2$, $f_{rep,i}$, $N_i$, and FSR$i$ denote the repetition rate, harmonic order, and free spectral range, respectively; $\Delta f_{rep}$ denotes the repetition-rate difference. Other symbols: $\nu$ and $f$, optical and radio frequencies, respectively; $\nu_0$, shared laser frequency (dashed line); $f_{RF}$, RF drive frequency; $\Delta\nu$, optical bandwidth.

## 3. Results

### 3.1. Experimental setup and comb characterization

Figure 2a illustrates the experimental setup of the gain-assisted EO comb. The output of a CW seed laser (ID Photonics CoBrite DX1, typical linewidth of 10 kHz) is coupled into a gain-assisted fiber ring cavity, in which the erbium-doped gain fiber is pumped by a laser diode through a wavelength-division multiplexer (WDM). The comb output is subsequently amplified by a fiber amplifier to an output power of 100 mW. The comb generates optical pulses at harmonics of the fundamental FSR of 197.06 MHz, which is determined by the RF signal generator. An extinction ratio exceeding 75 dB is achieved. For example, the measured RF spectrum exhibits a prominent peak at the modulation frequency ($f_{rep}$ = 4927.42 MHz; Figure 2b), with no observable signal near the fundamental FSR (Figure 2c). Notably, the comb phase noise, measured at $f_{rep}$, closely follows that of the RF signal source at offset frequencies below 100 kHz (Figure 2d). Above 100 kHz, additional noise contributions, including amplified spontaneous emission from the amplifier and electrical noise from the photodetector, elevate the noise floor [30, 31]. Nevertheless, the comb phase-noise spectral density approaches −135 dBc $Hz^{-1}$ at offset frequencies above 1 MHz.

The output spectrum of the comb exhibits a smooth spectral envelope (Figure 2e), as measured using an optical spectrum analyzer (Yokogawa AQ6370D) with a spectral resolution of 0.02 nm. The comb spectrum spans from 1555 to 1580 nm (with 15-nm or 1.8-THz bandwidth at -20 dB). The corresponding autocorrelation trace is shown in Figure 2f. The measured autocorrelation FWHM is 705 fs, close to the value of 633 fs calculated from the measured spectrum (Figure 2f). The ratio of 1.11 indicates that the output pulses are near transform-limited. We note that the comb power, pulse duration and spectral coverage can be further improved through additional amplification, pulse compression and nonlinear spectral broadening. For example, using a high-power fiber amplifier, we boost the comb power to 1.4 W and subsequently compress the pulses to 260 fs (Supplementary Figure S1). The amplified and compressed pulses are then launched into 1 m of highly nonlinear fiber (HNLF), generating a supercontinuum spanning approximately 1150–1950 nm (Supplementary Figure S1).

An important advantage of our system over conventional Fabry–Pérot EO combs [22] is its wide repetition-rate tunability without substantially compromising the comb spectral bandwidth. This tunability is achieved by selecting different harmonic orders of the cavity FSR; consequently, the $f_{rep}$ must be an integer multiple of the FSR. As shown in Figure 2g, decreasing $f_{rep}$ from approximately 5 to 0.2 GHz through successive harmonic orders results in only a gradual narrowing of the comb spectrum. Despite a 25-fold change in repetition rate, the 20-dB spectral bandwidth remains within 10–20 nm. Note that the current upper limit of the comb repetition rate (5 GHz) is imposed by the bandwidth of the signal generator used in our experiment. Additional examples are provided in Supplementary Figure S2. With a higher-bandwidth signal generator, the repetition rate could be further increased up to the bandwidth limit of the phase modulator (20 GHz).

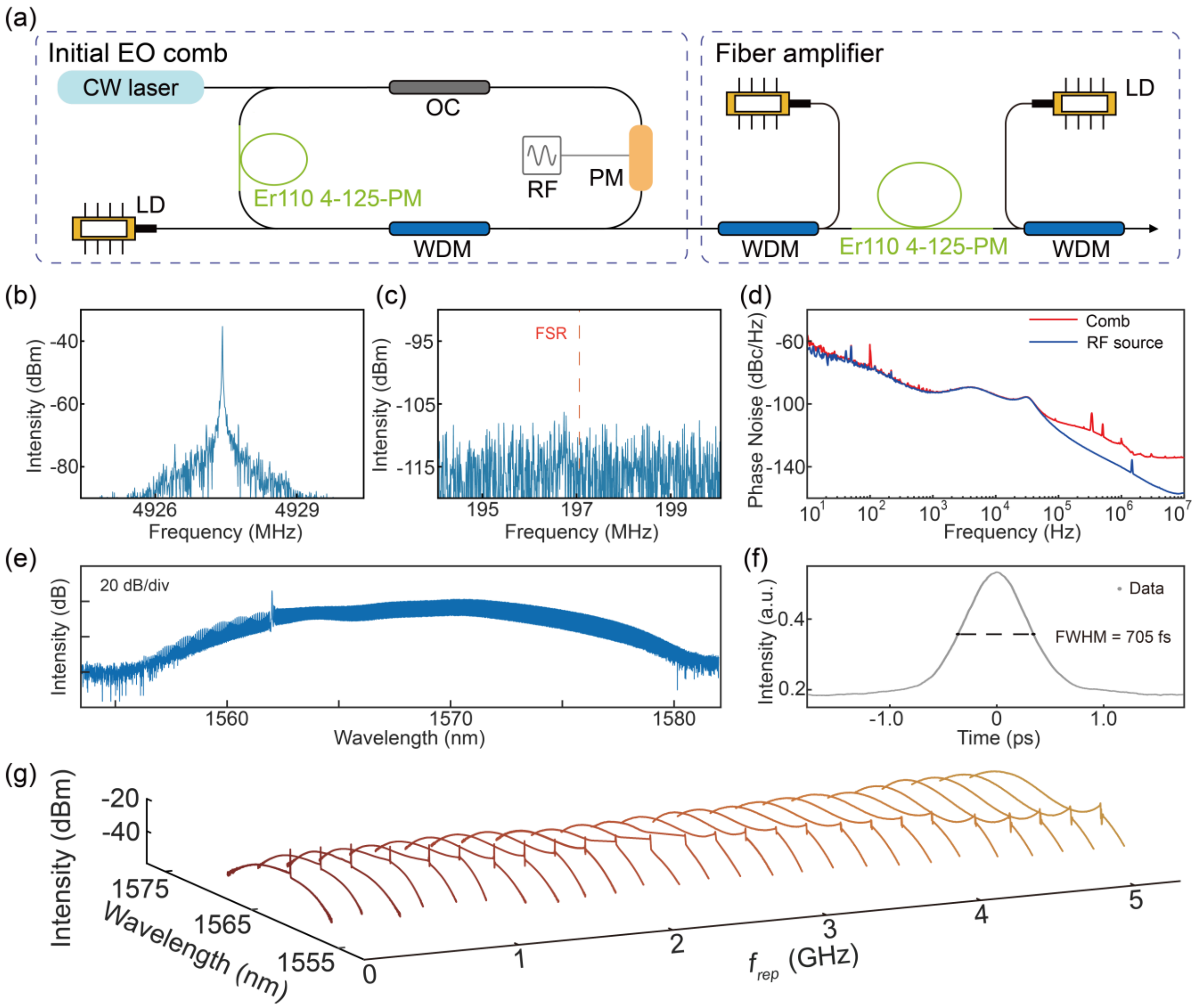

**Figure 2.** Experimental setup and optical comb characterization. (a) Electro-optic (EO) comb generation and fiber amplifier. (b,c) Radio-frequency (RF) spectra near the repetition rate ($f_{rep}$) and cavity free spectral range (FSR), respectively. The dashed line in (c) marks the FSR. RF resolution bandwidth (RBW) is 40 Hz. (d) Phase noise of the detected repetition-rate signal (red) and RF source (blue). The data are measured by a phase noise analyzer (FSWP8, Rohde & Schwarz). (e) Optical spectrum measured with a spectral resolution of 0.02 nm. (f) Intensity autocorrelation trace measured by an autocorrelator (pulseCheck SM 2000, APE). The dashed line in (f) indicates the full width at half maximum (FWHM). (g) Optical spectra measured at successive harmonic repetition rates. CW, continuous-wave; OC, optical coupler; PM, phase modulator; WDM, wavelength-division multiplexer; LD, laser diode.

### 3.2. Stability and operating tolerances

Next, we assess the spectral stability of a single comb resonator without active cavity feedback. As shown in Figure 3a, the fluctuations in the 20-dB spectral width remain below 1.3 nm, with a standard deviation of approximately 0.3 nm, corresponding to just 1.7 % of the nominal 20-dB spectral width (18 nm). Mechanical creep or temperature variation may alter the cavity optical path length, resulting in a mismatch between the cavity FSR and the RF driving frequency (i.e., $f_{rep}$). We should also note that touching the fiber cavity by hand can perturb the output spectrum. However, once the disturbance is removed, the comb spectrum gradually recovers, as illustrated by the spectral evolution beginning at 115 min in Figure 3a.

To characterize the spectral response to mismatches between the cavity FSR and $f_{rep}$, we record the comb spectra while scanning $f_{rep}$ around a resonant frequency at 5123.39 MHz in 0.01-MHz increments. The results are presented in Figure 3b. The comb exhibits a finite tolerance to FSR- $f_{rep}$ mismatch, arising from the interplay between intracavity nonlinearity and group-velocity dispersion. However, within a specific frequency range, indicated by the black arrow, the comb spectrum collapses due to self-excited oscillations [32].

Furthermore, we evaluate the comb's tolerance to temperature variations and mismatches between the cavity FSR and $f_{rep}$. Figure 3c shows the comb spectra measured at different operating temperatures. For this measurement, the comb source is placed in a temperature-controlled enclosure. At $f_{rep}$ = 5123.39 MHz, the spectral profile varies by less than 2 nm as the temperature increases from 22 to 25 °C, demonstrating relatively low sensitivity to temperature fluctuations over this range. Such a temperature range is readily accessible with conventional temperature controllers. Therefore, appropriate temperature regulation of the fiber can ensure good spectral stability of the comb source.

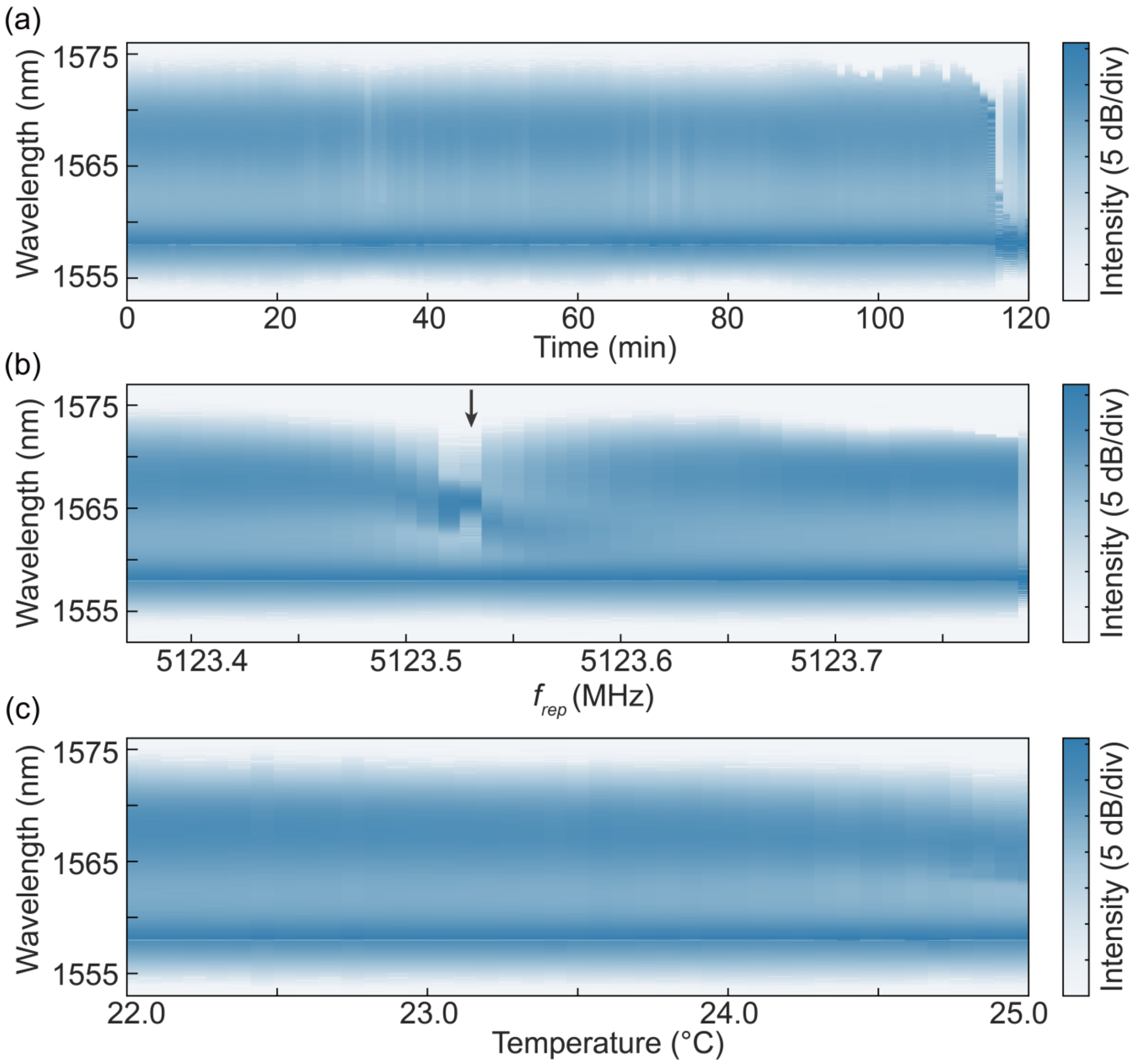

**Figure 3.** Spectral stability and operating tolerances of a single resonator. (a) Spectral evolution over 120 min, including recovery after a mechanical perturbation near 115 min. (b) Spectral evolution during a repetition-rate scan; the arrow marks the frequency range where the comb spectrum collapses. (c) Spectral evolution with cavity temperature. All spectra are measured before the fiber amplifier using an optical spectral analyzer (0.02 nm resolution). For (a) and (c), the repetition rate ($f_{rep}$) is fixed at 5123.39 MHz.

### 3.3. Dual-comb spectroscopy

Finally, we demonstrate the gain-assisted dual-resonator architecture for DCS measurement. The experimental setup is illustrated in Figure 1a. The two fiber-ring resonators share the same CW laser and are driven by two RF signal generators, both referenced to a commercial rubidium clock with a frequency stability of $10^{-11}$ at 1 s. The two resonators operate at repetition rates of 4926.34 and 4930.15 MHz, respectively, yielding a repetition-rate difference of $\Delta f_{rep}$ = 3.81 MHz. The two combs are heterodyned on a balanced photodetector (BPD), producing a time-domain interferogram with recurring bursts separated by approximately 0.26 μs (Figure 4a), determined by $1/\Delta f_{rep}$. For avoiding frequency aliasing, the two combs are spectrally filtered [4]. The dual-comb signals are recorded in the time domain using an oscilloscope (DSAX93204A, Agilent) with a 1-GHz bandwidth and a sampling rate of 2.5 GS/s. The corresponding RF spectrum (Figure 4b) exhibits regularly spaced multiheterodyne beat notes spanning approximately 1 GHz.

The two combs exhibit excellent mutual coherence. For instance, the dual-comb beat notes, exemplified at 99.06 and 400.05 MHz (Figures 4c and 4d), have linewidths of approximately 2 Hz, reaching the Fourier-transform limit. These results demonstrate that the dual combs maintain mutual coherence for nearly 1 s without active frequency stabilization, providing sufficient coherence for a broad range of spectroscopic sensing applications. As an example, we measure the transmission spectrum of a fiber Bragg grating (FBG) using an asymmetric dual-comb architecture [4]. The measured spectrum, shown in Figure 4e, is comb-line-resolved and acquired in just 2 μs. The blue dots spaced by 4.9 GHz or 40 pm indicate the peak amplitudes of individual comb lines. The retrieved transmission profile agrees well with the reference spectrum (resolution of 40 pm) measured using an optical spectrum analyzer, which requires an acquisition time of 0.01 s. Our system enables rapid, continuous, single-shot monitoring of FBG transmission spectra. An example is shown in Figure 4f, where each spectrum is acquired within 262.47 ns, corresponding to one dual-comb cycle.

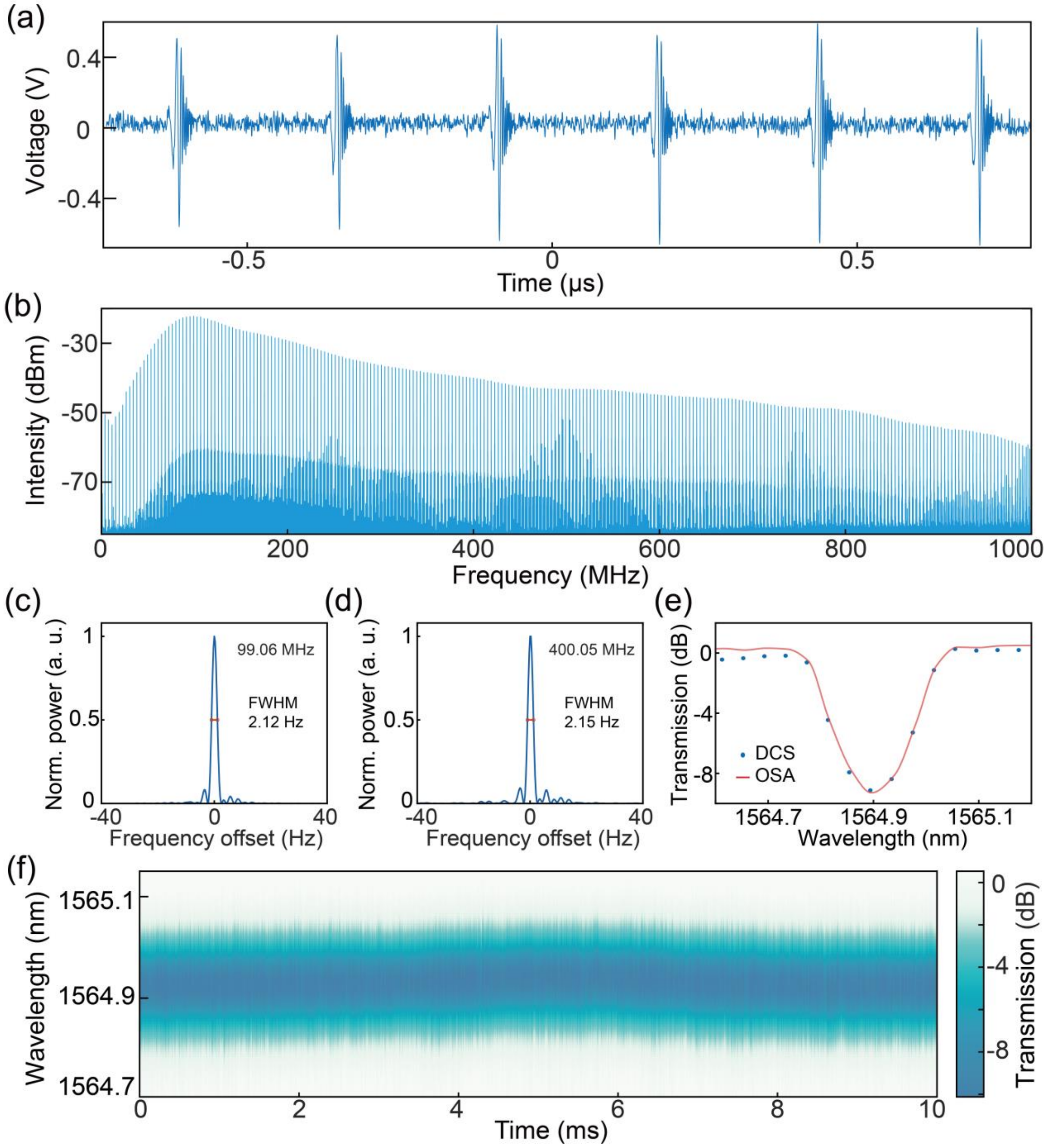

**Figure 4.** Dual-comb results. (a) Time-domain interferogram. (b) Radio-frequency (RF) multiheterodyne spectrum. (c,d) Normalized beat-note spectra near 99.06 and 400.05 MHz, respectively. Frequency offsets are referenced to the line centers, and red circles mark the half-maximum points. (e) Fiber Bragg grating (FBG) transmission measured by dual-comb spectroscopy (DCS; blue points) and an optical spectrum analyzer (OSA; red line). (f) Temporal evolution of the FBG transmission response.

## 4. Discussion

Our work presents a simple approach to broadband dual-comb generation with widely tunable repetition rates and coherent multiheterodyne detection, without the need for complex frequency stabilization. Table 1 compares our approach with previously reported EO comb systems in terms of their 20-dB bandwidths, numbers of comb lines and repetition rates. Note that this comparison excludes systems employing additional nonlinear media for spectral broadening, as such nonlinear processes can generally degrade comb coherence [33, 34]. We note that the spectral bandwidth of the fiber gain medium is a key factor limiting the spectral coverage of the generated combs. Extending the gain bandwidth through alternative amplification schemes, such as Raman amplification [35] or optical parametric amplification [36], could enable dual-comb operation in other spectral regions of interest [19, 36].

**Table 1.** Comparison of broadband electro-optic comb systems.

| **Generation scheme** | $f_{\mathrm{rep}}$ **[GHz]** | $\Delta\nu_{20\ \mathrm{dB}}$ **[THz]** | $N_{20\ \mathrm{dB}}$ | **Ref.** |
|---|---|---|---|---|
| Single-pass cascaded intensity modulator (IM) and three PMs | 10 | 0.67 | 70 | [21] |
| Single-pass cascaded IM and three PMs | 25 | 3.3 | 130 | [37] |
| Fiber loop with dispersion compensation | 19.8 | 3.1 | 160 | [28] |
| Coupled microring resonators | 30.9 | 6.3 | 200 | [24] |
| Micro-resonator with multi-tone driving | 4.2 | 0.88 | 210 | [38] |
| Microring resonator | 10.4 | 4.4 | 420 | [39] |
| Triply resonant comb generator | 29.6 | 36 | 1200 | [25] |
| Gain-assisted fiber ring with a single PM and dispersion management | 1<br>5 | 1.4<br>1.8 | 1400<br>360 | This work |

Our all-fiber system offers considerable potential for a broad range of applications requiring flexible, broadband and high-speed precision measurements. The widely tunable repetition rates enable flexible adjustment of the spectral resolution and measurement bandwidth, making the system adaptable to different spectroscopic requirements. Potential applications include high-resolution molecular spectroscopy [7, 15], time-resolved and transient spectroscopy [40], trace-gas sensing [8], and precision frequency-domain measurements [15, 41]. In particular, the spectral coverage over 1 THz is advantageous for simultaneously probing multiple molecular absorption features, enabling the identification and quantitative analysis of complex gas mixtures [8]. The coherent multiheterodyne detection further provides a direct and efficient means of retrieving spectral information with high measurement precision, which is beneficial for monitoring fast dynamic processes [42] and capturing transient phenomena [40]. Beyond spectroscopy, the flexible repetition-rate tuning may benefit applications in optical metrology, frequency calibration and precision distance measurements [43, 44]. Moreover, the all-fiber configuration, together with its simplified operation and reduced requirements for complex frequency stabilization, offers practical advantages in terms of compactness and ease of integration. These features may facilitate the development of portable and cost-effective dual-comb instruments for field-deployable chemical sensing [45], environmental monitoring [3], industrial process analysis [46] and other applications where reliable broadband measurements are required outside laboratory environments.

## 5. Conclusion

In conclusion, we demonstrate an all-fiber, gain-assisted dual-comb architecture for broadband (>1 THz) coherent multiheterodyne detection. The two combs exhibit excellent mutual coherence, with a relative linewidth of 2 Hz, without requiring active frequency stabilization. Moreover, their repetition rates can be widely tuned from 0.2 to 5 GHz while maintaining broadband spectral coverage. This combination of broad bandwidth, flexible repetition-rate tuning, high mutual coherence and simplified operation makes our approach a promising platform for dual-comb spectroscopy and a wide range of precision spectroscopic sensing applications.

## Abbreviations

BPD, balanced photodetector; CW, continuous-wave; DCS, dual-comb spectroscopy; EDF, erbium-doped fiber; EO, electro-optic; FBG, fiber Bragg grating; FSR, free spectral range; FWHM, full width at half maximum; HNLF, highly nonlinear fiber; IM, intensity modulator; InP, indium phosphide; ISO, optical isolator; LD, laser diode; LN, lithium niobate; LT, lithium tantalate; OC, optical coupler; OSA, optical spectrum analyzer; PD, photodetector; PM, phase modulator; RBW, resolution bandwidth; RF, radio-frequency; WDM, wavelength-division multiplexer.

## Supplementary Information

Additional file 1: Supplementary Information.docx. Figure S1: Power amplification and spectral broadening. Figure S2: Comb spectra at different repetition rates.

## Declarations

## Availability of data and materials

The data that support the findings of this study are available from the corresponding author upon reasonable request.

## Competing interests

The authors declare no competing interests.

## Funding

This work is funded in part by the Innovation Program for Quantum Science and Technology (No. 2023ZD0301000).